\documentclass[conference]{IEEEtran}
\usepackage[utf8]{inputenc}
\usepackage[frozencache,cachedir=.]{minted}

\IEEEtriggeratref{21}

\usepackage{cite}
\usepackage{amsmath,amssymb,amsfonts}
\usepackage{algorithmic}
\usepackage{graphicx}
\usepackage{textcomp}
\usepackage{xcolor}
\usepackage{url}
\usepackage[normalem]{ulem}

\usepackage{boxedminipage}
\newenvironment{result}
{\smallskip
\noindent
\let\emph=\textbf
\begin{boxedminipage}{\columnwidth}\em}
{\end{boxedminipage}
}

\begin{document}

\title{Model-based Automated Testing of Mobile Applications: An Industrial Case Study}

\author{
\IEEEauthorblockN{Stefan Karlsson\IEEEauthorrefmark{1}\IEEEauthorrefmark{2}, Adnan \v{C}au\v{s}evi\'{c}\IEEEauthorrefmark{1}\IEEEauthorrefmark{2}, Daniel Sundmark\IEEEauthorrefmark{1},
Mårten Larsson\IEEEauthorrefmark{2}
}

\IEEEauthorblockA{\IEEEauthorrefmark{1} Mälardalen University, Västerås, Sweden}
\{stefan.l.karlsson,daniel.sundmark\}@mdh.se
\IEEEauthorblockA{\IEEEauthorrefmark{2} ABB AB, Västerås, Sweden}
\{stefan.l.karlsson,adnan.causevic1,marten.larsson\}@se.abb.com
}

\maketitle

\begin{abstract}
Automatic testing of mobile applications has been a well-researched area in recent years. However, testing in industry is still a very manual practice, as research results have not been fully transferred and adopted. Considering mobile applications, manual testing has the additional burden of adequate testing posed by a large number of available devices and different configurations, as well as the maintenance and setup of such devices.

In this paper, we propose and evaluate the use of a model-based test generation approach, where generated tests are executed on a set of cloud-hosted real mobile devices. By using a model-based approach we generate dynamic, less brittle, and implementation simple test cases. The test execution on multiple real devices with different configurations increase the confidence in the implementation of the system under test. Our evaluation shows that the used approach produces a high coverage of the parts of the application related to user interactions. Nevertheless, the inclusion of external services in test generation is required in order to additionally increase the coverage of the complete application. Furthermore, we present the lessons learned while transferring and implementing this approach in an industrial context and applying it to the real product.
\end{abstract}{}


\maketitle

\section{Introduction}

Automated testing of a graphical user interface (GUI) of a software application is commonly done by manually produced test cases, a process that can result in brittle and static tests. The brittleness of GUI tests has been related to the methods of finding, enumerating and evaluating the state of specific UI elements, but also to the fact that a UI is a stateful process were current navigation possibilities and actions depend on those previously executed. This results in multiple test cases executing parts of the same navigation path of the GUI to reach different end-states. If a navigation property is changed, such as the addition of a new screen, all tests on a path related to the new screen must be updated accordingly. 

Cost-benefit comparison of manually written tests to automatically generated tests is seen as a trade-off of the time a human must spend on writing tests in contrast to producing the input assets required by an automatic method. In addition to time spent, there is also the dimension of how static the test cases are. Manually written tests will execute the same test cases until manually changed, while, depending on the method used, automatically produced tests have the potential to create new test cases. Test generation for a GUI thus has the potential of increasing the coverage of the vast state space of the possible stateful interactions. 

Considering mobile applications as a specific case of a GUI application we get the added challenges and complexity, both in development and testing, of the availability and differences in mobile devices and platforms where the application can be deployed \cite{Real-Challenges-in-Mobile-App-Development}. Different mobile devices have different operating system (OS) versions, screen sizes, screen resolutions, etc. that can affect both the visual appearance of the application but also its operational properties as CPU and memory usage. In addition, measuring and analyzing metrics of mobile devices, such as CPU, memory and battery usage, can be a challenge~\cite{Real-Challenges-in-Mobile-App-Development}.

To address the issues of test cases being static, only covering a small part of the possible state space, test case maintenance due to changes in the GUI, and the configuration differences in used devices, we propose an approach that combines model-based testing, random exploration of the model, and system under test (SUT) execution in cloud-hosted mobile devices. This approach takes a model of the GUI interactions as input and produces a random exploration covering all the states in the model. This is further executed on a fleet of cloud-hosted devices while logging performance metrics. In addition to the model serving as input to the test generation process, it also serves as a documentation of the states of the application. Updating the model then keeps the documentation and tests current and in sync.

Research in mobile application testing, specifically on the Android platform, has been an active area in recent years \cite{Kong-Automated-Testing-of-Android-Apps-review, A-systematic-mapping-study-of-mobile-application-testing-techniques}, resulting in tools such as MobiGUITAR \cite{MobiGUITAR}, SAPIENZ \cite{Sapienz} and APE \cite{Practical-GUI-Testing-of-Android-Applications}. However, manual testing is still a large part of testing mobile applications \cite{Understanding-the-Test-Automation-Culture-of-App-Developers, Real-Challenges-in-Mobile-App-Development}, for several reasons, such as time constraints but also due to the lack of knowledge of the available tools and methods \cite{Understanding-the-Test-Automation-Culture-of-App-Developers}. This is also our experience from industry where automation is considering only the \textit{execution} of manually written tests, not the \textit{generation} of tests. Thus results from academia have not yet been transferred to industry~\cite{how-do-developers-test-android} and more evaluations in real-word contexts are needed~\cite{A-systematic-mapping-study-of-mobile-application-testing-techniques}, which we have aimed to do with this work. 

In this paper, we contribute with a model-based approach and implementation of testing a GUI application, deployed to multiple cloud-hosted devices, providing metrics of CPU and memory usage. Additionally, we provide evaluation of using the method in an industrial use-case at our industry partner, ABB. The results of our evaluation show that the effectiveness of the approach, when covering functionality related to user interactions of the SUT, is high. However, there is room for improvement to increase the coverage related to the handling of externalities such as service communication and authorization. Finally, we discuss the lessons learned of transferring knowledge of model-based testing from academia to industry.

\section{Motivating Example}

A large part of mobile application testing activities in industry are still manual practices \cite{Understanding-the-Test-Automation-Culture-of-App-Developers, Real-Challenges-in-Mobile-App-Development}. However, when our industry partner, ABB, was developing a new mobile application, targeting Android and iOS, an effort to ensure quality by using automatic tests was considered. Based on previous experiences of test automation efforts, it has been hard to motivate developers to reach a high level of test automation. In addition, the perception from developers, with regards to test automation of GUI applications, is that they are more brittle than tests of lower-level functionality. Further, targeting mobile devices, there is a wast set of possible device configurations to verify. Maintaining such a large set of devices and configuration in-house was deemed as not feasible. Manually written tests are also seen as a collection of static examples and to find more complex bugs, a method of producing more dynamic tests was of high interest. In summary, the problems of automatic testing for the organization are:
\begin{itemize}
    \item Automatic tests are lacking or have low coverage.
    \item Automatic tests are perceived as brittle.
    \item Automatic tests are perceived as static, not finding more complex interactions.
    \item There is a need for a feasible strategy to handle a high number of different device configurations.
\end{itemize}{}

The objective of this work was thus to find a simple and practical approach that could be adopted by developers in a short amount of time, with the aim to increase the coverage of automated tests, reduce the brittleness of tests, reduce the developer burden of writing tests by leveraging test generation, and allow for more dynamic tests.

In a survey of Android developers, it was found that developers rely on high-level usage models, such as use-cases or requirements, to model the test cases for applications. In addition, developers seem to prefer test-cases focused around high-level concepts, such as features. Further, the survey shows that tests are usually manually written and using GUI automation tools, resulting in fragile tests that break with changes to the GUI. High-level coverage, as use-case coverage or feature coverage, were seen as more valuable than code coverage \cite{how-do-developers-test-android}. We have considered an approach that solves the use-case of our industry partner, but in doing so, we also perceive the approach to be a good fit for the generally reported developer preferences.

\section{Used approach}\label{section:proposed-method}
\begin{figure*}[h]
    \includegraphics[width=\textwidth]{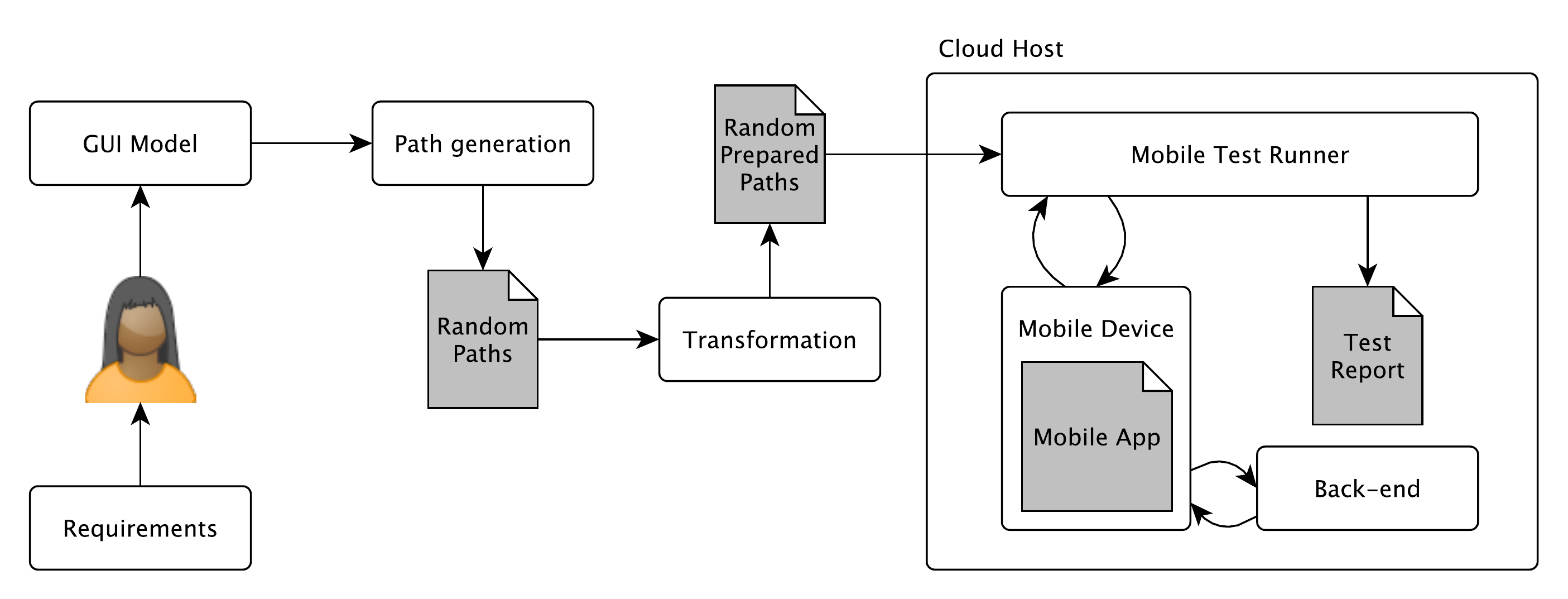}
    \caption{Overview of the test case generation method}
    \label{fig:method-overview}
\end{figure*}

\begin{figure*}[]
    \includegraphics[width=\textwidth]{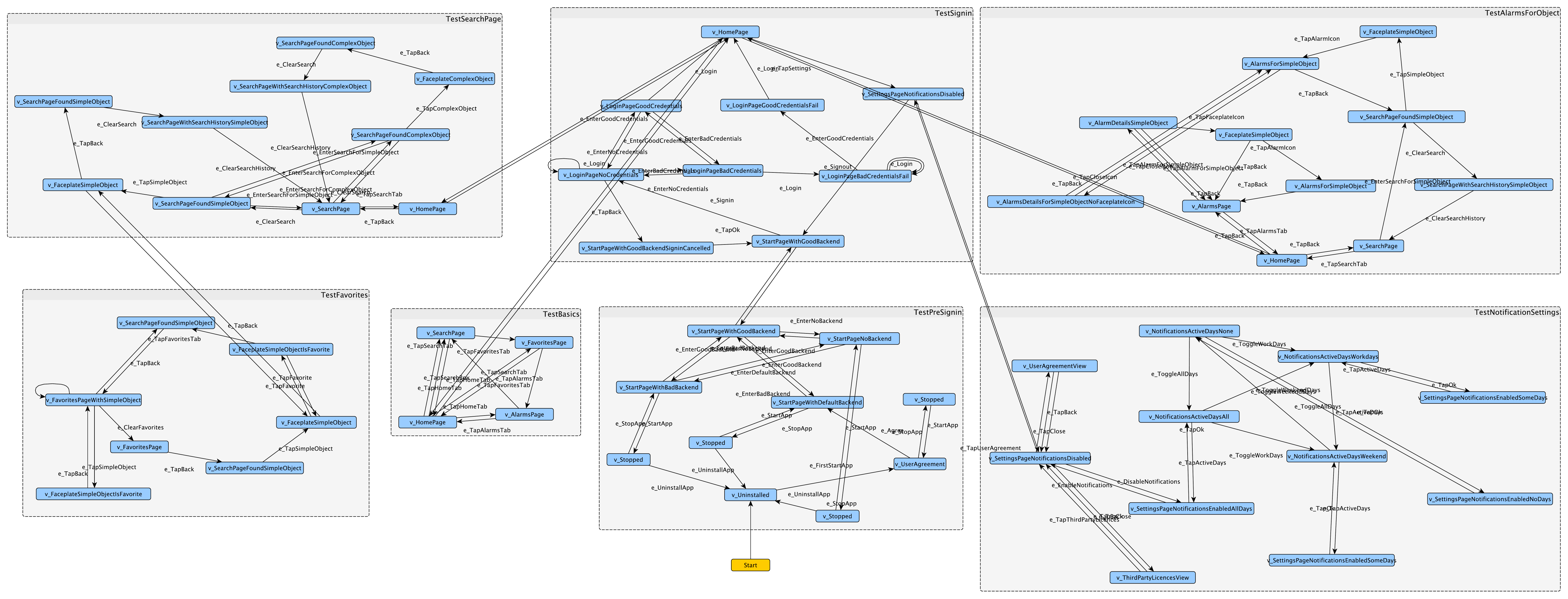}
    \caption{Overview of the model used. Each gray box represents a feature of the application.}
    \label{fig:implementaiton-model}
\end{figure*}

\begin{figure}[]
    \includegraphics[width=0.5\textwidth]{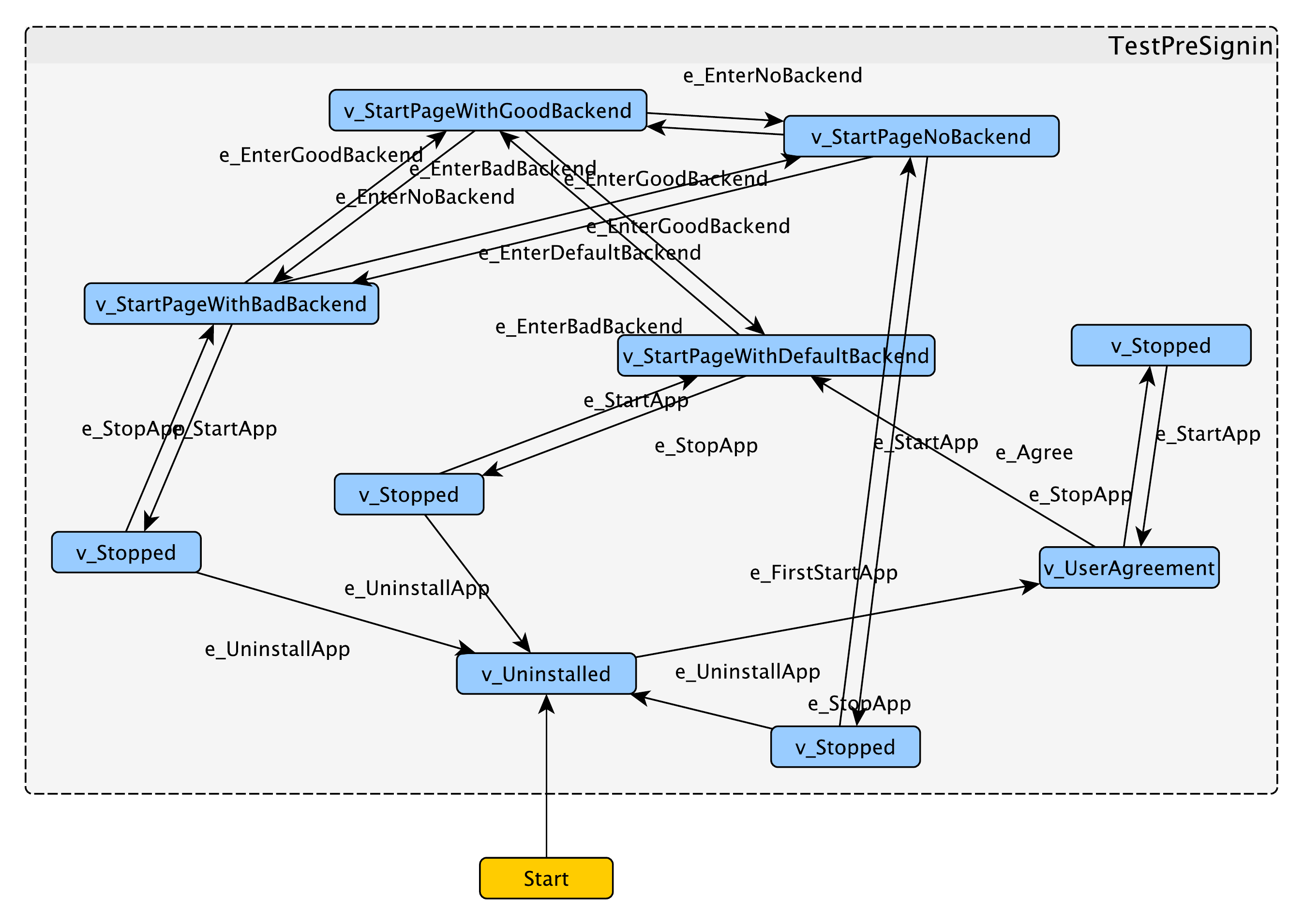}
    \caption{Feature subgraph}
    \label{fig:implementaiton-test-case}
\end{figure}

In this section, we start by providing an overview of all the steps used in the approach, prior to giving a more in-depth explanation of the individual activities. Given the taxonomy of model-based testing approaches proposed by Utting et al. \cite{A-taxonomy-of-model-based-testing-approaches}, we have used an event-discrete, transition-based model with search-based test generation, targeting structural model coverage, in an offline fashion, i.e., test generation is separated from test execution.

Figure \ref{fig:method-overview} shows an overview of the complete process, including model creation, test generation and test execution on a cloud-hosted set of devices. The steps included are:

\begin{enumerate}
    \item Producing a model describing the usage of the GUI of the App from requirements.
    \item Generating a sequence of node and edge transitions.
    \item Transforming the sequence to a suitable format for test execution.
    \item Executing the generated path by Test Runner.
    \item Producing test report.
\end{enumerate}

\subsection{The model}
In our approach, we used a manually created model on a high abstraction level. The model represents the application from the perspective of a user, where nodes represent user-perceived states and edges represent actions performed by the user. In this sense, the model is a black-box model of the application under test. It models the features of the application, but it does not depend on its implementation. In addition to modeling the states representing features of the application, the model also includes the user's observation of the device, where the application will run, such as if the application is started, stopped, installed or uninstalled. Including device level states allow for test generation to include sequences were different behavior of the application is expected in case the application has been previously started or not.

Research often proposes approaches that aim to automatically create a model of the application under test and later use it to generate test cases. In our approach, we have opted to use a manually created model. There are several trade-offs to consider when deciding on a manual or automatic approach for model creation. An automatic approach has the benefit of efficiency, generating models without requiring human interaction, and generalization, modeling of a large set of applications. This can be an attractive proposition for researchers who want to apply and evaluate a method on a large range of applications. However, for development teams in industry, who might develop and maintain a low number of applications, this is less of a benefit. Instead, in our experience, it is more valuable for practitioners in industry to have a model that correctly specifies the specific behavior of the application they currently develop and express it on an abstraction level of the \textit{user} rather than of the \textit{user interface elements}. Keeping a high abstraction level reduces the state space explosion possible by using a GUI element based model, with the additional benefits of the model serving as a design and documentation specification of the application being developed.

The definition of the model, $M$, is defined as $M=(N,E)$ where $N$ is the set of \textit{nodes}, $N=\{n_0, n_1, ..., n_n\}$, and $E$ the set of \textit{edges}, $E=\{e_1, e_2, ..., e_n\}$. The nodes represent the states of the application and the edges represents user actions, transitioning the application between its different states.

\subsection{Path and Walk Generation}\label{section:generation}
We generate both \textit{paths}, were nodes are distinct and \textit{walks}, were the same node can be contained several times, when generating test cases and combine those into a test case sequence. We facilitate two scenarios for test generation: (i) considering the complete model for test generation, or (ii) considering only a subgraph of the model. Figure \ref{fig:implementaiton-model} shows a complete model where each box contains a subgraph representing a feature in the SUT. Figure \ref{fig:implementaiton-test-case} shows one such subgraph. The reason for making this division of the model is to allow for test case generation of features as individual test cases.

Considering that the navigation of the SUT is stateful, and to be able to test a subgraph, we must reach the starting state of the subgraph, i.e., the initial state of the feature to be tested from the starting node of the model. Looking at the model in Figure \ref{fig:implementaiton-model}, we can see that to test the feature in the top left box we must first navigate through the graphs in the middle two boxes. In practice, this can mean that to be able to start executing the tests for a "search-page" we must first go through "sign-in" and "navigate to search" subgraphs.

To be able to run several tests in a sequence without having to reset the test environment we want the test cases to end in a known navigation state. This allows the next test case to start from a known node in the graph. Thus, we want to generate a path from the final node of a generated test case to the shared known starting node for several feature subgraphs. In practice, this might be a common "Home page" of the application.

This results in three sets of generated nodes, two paths, and one walk. The first path, $P_{\mathit{setup}}$, is the path from the starting node of the model to the starting node of the subgraph to be tested, a random walk, $W_{\mathit{random}}$, of the subgraph under test and the path to the final node, $P_{\mathit{teardown}}$. We end up with the following definitions:
Given that $n_0$ is the starting node of the model $M$, $s_0$ is the starting node of subgraph $S$, and $s_f$ is the final state of $S$, then

\begin{itemize}
    \item $\{n_0, s_0, s_f\} \subseteq M$
    \item $P_{\mathit{setup}} = [n_0, n_i, ..., s_0]$
    \item $W_{\mathit{random}} = [s_0, ..., s_i]$
    \item $P_{\mathit{teardown}} = [s_i, ..., s_f]$
    \item $Test = [P_{\mathit{setup}}, W_{\mathit{random}}, P_{\mathit{teardown}}]$
\end{itemize}{}

The stopping criteria used in the random walk of nodes and edges in $W_{random}$ is to cover all edges in the selected subgraph. This is done by randomly selecting an edge in the subgraph not visited, generate a path to the edge, mark edges in the path as visited and repeat until all edges in the subgraph have been visited. Since nodes are allowed to have multiple edges we use an edge based stopping criteria instead of a node covering criteria. An edge covering criteria will contain all nodes, but the opposite is not necessarily true.

\subsection{Sequence Transformation}
The generated sequence of transitions is expressed in the abstract format of nodes and edges of the model. In the transformation step, we transform the abstract sequence to a format usable by the specific test runner. In practice, this turns out to be a straightforward step since we can map nodes and edges to function names available to the test runner. The test runner implementation will thus contain implementations of all the unique nodes and edges of the model. Typically the function of an edge will be to perform an action, transitioning the application to a new state, and the function of nodes will be to assert that the state of the application is as expected.

The method of mapping nodes and edges to functions allows for a high level of reuse and simplicity of the test runner. Common transition actions will share the same implementation and the implementation of nodes are simple since it is a verification of the state at the current node. Thus, we do not need any other context than the current application state. This method also solves a lot of the perceived brittleness of testing on the user level since changes in the model only require isolated changes in the specific nodes and edges, the implementation is agnostic to the composition of the edges and nodes used. In addition, with a growing set of node and edge implementations, the need for new implementation effort diminishes.

\subsection{Test Execution}
The execution of the prepared test sequence is performed by a test runner which is adapted to the platform running the application under test, e.g., mobile application, web application, etc. To be able to test a SUT given multiple device configurations, such as mobile devices or different web-browsers, we propose to execute the tests in a cloud-hosted device farm. The reason for using a cloud-based solution is to allow developers to select a relevant set of devices and configurations, that are applicable for the current use-case, from a larger set of available devices. In this fashion, the configuration and hosting of the devices are abstracted for the user.

The cloud host will provide a test report containing the pass/fail result of the executed test cases and, in addition, the report will contain performance counters such as CPU and memory usage during the course of the test execution. The test report also contains optional screenshots taken during the test execution. This allows developers to view the actual GUI at selected test execution steps. The ability to view screenshots is especially valuable considering different devices and different screen orientations and resolutions.
\section{Implementation}\label{section:implementation}

\begin{figure}
    \centering
\begin{minted}{json}
{"currentElementName":"v_StartPageWithGoodBackend"}
{"currentElementName":"e_Signin"}
{"currentElementName":"v_LoginPageNoCredentials"}
{"currentElementName":"e_EnterBadCredentials"}
{"currentElementName":"v_LoginPageBadCredentials"}
{"currentElementName":"e_Login"}
{"currentElementName":"v_LoginPageBadCredentialsFail"}
{"currentElementName":"e_EnterGoodCredentials"}
{"currentElementName":"v_LoginPageGoodCredentialsFail"}
{"currentElementName":"e_Login"}
{"currentElementName":"v_HomePage"}
{"currentElementName":"e_TapSettings"}
{"currentElementName":"v_SettingsPageNotificationsDis.."}
{"currentElementName":"e_Signout"}
{"currentElementName":"v_StartPageWithGoodBackend"}
{"currentElementName":"e_Signin"}
{"currentElementName":"v_LoginPageNoCredentials"}
\end{minted}
    \caption{GraphWalker output example of a Quick Random walk of one feature subgraph in JSON}
    \label{fig:gw-output}
\end{figure}

\begin{figure}
    \centering
\begin{minted}{bash}
v_StartPageWithGoodBackend
e_Signin
v_LoginPageNoCredentials
e_EnterBadCredentials
v_LoginPageBadCredentials
e_Login
v_LoginPageBadCredentialsFail
e_EnterGoodCredentials
v_LoginPageGoodCredentialsFail
e_Login
v_HomePage
e_TapSettings
v_SettingsPageNotificationsDisabled
e_Signout
v_StartPageWithGoodBackend
e_Signin
v_LoginPageNoCredentials
\end{minted}
    \caption{GraphWalker transformed output}
    \label{fig:gw-transformed}
\end{figure}

In this section, we describe the specific implementation of the generic approach presented in Section \ref{section:proposed-method}. Some of the implementation parts are more generic than others. For example, GraphWalker is highly re-usable for any type of SUT but the implementation of the test runner targeting a mobile application user interface is highly specific to the target technology.

It turns out that many of the parts needed to implement the approach can be done by leveraging off-the-shelf tools and components, combined with some implementation given the specific use-case. The toolbox used in our implementation consists of:

\begin{itemize}
    \item yEd\footnote{\url{https://www.yworks.com/products/yed}}
    \item GraphWalker\footnote{\url{http://graphwalker.github.io/}}
    \item GNU Bash\footnote{\url{https://www.gnu.org/software/bash/}} scripts
    \item Xamarin.UITest\footnote{\url{https://docs.microsoft.com/en-us/appcenter/test-cloud/uitest/}}
    \item App Center\footnote{\url{https://appcenter.ms/apps}}
\end{itemize}

In addition to being able to implement parts of the approach, tools have also been selected on their easy-of-use and barrier-of-entry, since, in our experience, these factors are of high importance when targeting industry adoption.

\subsection{Modeling with yEd}
yEd is a diagramming application, available as freeware. The application is written in Java, thus making it available for all major platforms. yEd gives practitioners a simple and easy way of creating models in a "What You See Is What You Get"\footnote{\url{https://en.wikipedia.org/wiki/WYSIWYG}} fashion. In addition, a strong argument for yEd is that the diagram format, GraphML\footnote{\url{https://en.wikipedia.org/wiki/GraphML}}, used to save and load graphs by yEd, is readable by GraphWalker. 

In GraphWalker major version 4, GraphWalker provides a modeling capability in the form of GraphWalker Studio. However, at the start of this project, this version was only available in beta and thus the stable 3.x version was used. Consequently, we have not accessed the build-in Studio.

\subsection{Generation with GraphWalker}
GraphWalker is a model-based testing tool, available as open-source software \cite{graphwalker}. GraphWalker has the capabilities of generating paths and walks of directed graph models. GraphWalker can be used as a Java library, integrated into an application or test suite, a command-line tool or as a server. The GraphWalker project is currently, and has been for many years, an active open-source project. From that it is reasonable to infer that the tool is used by practitioners. In the literature GraphWalker is reported to have been used by Spotify\footnote{\url{https://www.spotify.com}} ~\cite{visual-gui-testing-in-practice} and in the Automotive Industry ~\cite{Automated-Model-Based-Software-Test-Cases-in-the-Automotive-Industry}.

GraphWalker supports several different generator algorithms, how the graph will be traversed, and coverage criteria, when GraphWalker will stop generation\footnote{\url{https://github.com/GraphWalker/graphwalker-project/wiki/Generators-and-stop-conditions}}. The interesting ones to our  approach (GraphWalker internal name in parenthesis) are; \textit{A*} (a\_star), \textit{Quick Random} (quick\_random) and \textit{Edge Coverage} (edge\_coverage). \textit{A*} is the commonly known search algorithm that will find the shortest path between two nodes or edges in the graph. We use \textit{A*} to generate the path of the test setup, $P_{\mathit{setup}}$, when testing features in a subgraph of the model,  as described in Section \ref{section:generation}. As also described in the approach we want a path from the last node of the feature subgraph to a commonly shared node for the test teardown, $P_{\mathit{teardown}}$, to serve as the starting node for the next test, increasing the efficiency of starting the next test case. As for the test setup, \textit{A*} could also be used for the teardown. However, due to limitations in the GraphWalker API, where \textit{A*} can only generate a path from a special \textit{Start node} to any other node, and not between any two nodes, the current implementation does not use \textit{A*} for the test teardown. The consequence of this limitation is that when a new random path is generated for a subgraph, the teardown sequence must manually be defined. Our industry partner sees the benefit in extending the implementation to also use \textit{A*} for the teardown and will do so in the next version of the implementation. This combination will give us the shortest path from the start of the model to the start of the subgraph feature under test and the path from the last visited node in the subgraph to its designated final node.  

\textit{Quick Random} is used to generate a random walk of the subgraph under test, $W_{\mathit{random}}$. We use 100\% edge coverage as a stopping criterion of \textit{Quick Random}. The difference between \textit{Random} and \textit{Quick Random} is that while \textit{Random} randomly selects an outgoing edge from the current node, \textit{Quick Random} selects an edge not visited and uses the shortest path to get there, adding the newly visited edge to the set of visited edges. Both algorithms are useful for random testing, while \textit{Quick Random} can be more suitable for large models, typically generating shorter paths.

The output from GraphWalker is a sequence of edge and node transitions. This sequence is based on the model and agnostic of implementation details of the SUT. It is thus up to developers and testers to implement the actions and assertions corresponding to the edges and nodes of the model, to enable execution of any generated sequence.

\subsection{Transformation with Bash}
When using GraphWalker in offline mode, i.e, generating sequences without the SUT running, we get a textual representation of the generated sequence, by default in JSON\footnote{\url{https://en.wikipedia.org/wiki/JSON}} format. Figure \ref{fig:gw-output} shows a partial example output generated from a \textit{Quick Random} walk of a subgraph. We then need to transform the GraphWalker representation to a format that makes sense to our test runner that will execute the tests. In our implementation, we made this transformation very straight forward by mapping the name of an edge or node to a method name in the test runner. Thus the output from GraphWalker of \texttt{\{"currentElementName":"e\_Login"\}} will be transformed to \texttt{e\_login}, which will be the expected name of a method in the test runner. This relationship can also be seen in Figure \ref{fig:gw-output} in relation to the transformed output in Figure \ref{fig:gw-transformed}.

\subsection{Test execution}
\begin{figure}
    \centering
\begin{minted}{csharp}
public static object 
  v_AlarmsForSimpleObject(object context) =>
    AppWaitForElements(
        context,
        ids: new string[]
        {
            // All these ids
            "detail_activity_menu_share",
            "detail_activity_menu_favorite",
            "master_alarm_row_condition",
        },
        texts: new string[]
        {
            // All these texts
            "Alarms",
        },
        idTexts: new (string, string[])[]
        {
            // All these ids
            ("master_alarm_row_state", new string[]
            {
                // Any of these texts
                "Active",
                "Inactive"
            }),
        });
\end{minted}
    \caption{Implementation of a node, which will assert on the current state}
    \label{fig:node-implementation}
\end{figure}

\begin{figure}
    \centering
\begin{minted}{csharp}
public static object 
  e_EnterSearchForSimpleObject(object context)
{
    AppEnterTextById(
        context, "search_src_text", SimpleObject);
    return context;
}
\end{minted}
    \caption{Implementation of an edge, which will perform an action on the SUT}
    \label{fig:edge-implementation}
\end{figure}

This part of the implementation is the most specific part since it is tied to the platform of where the SUT executes. In our implementation examples we discuss Xamarin.UITest, targeting a mobile application, but a similar implementation strategy can be used for a web-based application with, for example, the web-automation library Selenium\footnote{\url{https://www.selenium.dev/}}. 

As explained, edges and nodes in the model will have a one-to-one mapping to methods in the test runner. This approach keeps the implementation simple and straight forward as well as highly reusable. For example, in a mobile application the action to "tap back" is used frequently to navigate in the GUI. The implementation of such an action can then be reused throughout the model.

Figure \ref{fig:node-implementation} lists an example of a node implementation that will make an assertion based on the elements present in the GUI. In line 1 we can see the name of the method which maps to the name of the node in the model. The arguments for the \texttt{AppWaitForElements} method are the element ids expected (4-10), the texts expected (11-15) and the text ids expected (16-25). In Figure \ref{fig:edge-implementation} we can see an example of an edge implementation, an action to transition the application to the next state. As for the node implementation, we can see in the first line that the name of the method corresponds to an edge name of the model. The simple task of this edge is to enter a search string (3), given by a variable, into an element with a specific id.

The task of the test runner is thus to execute the generated sequence from the model by executing the corresponding edge and node implementations, transitioning in the edges and asserting in the nodes. Any assertion failure in a node will result in a failed test case.

\subsection{Cloud hosting with App Center}
\begin{figure*}[]
    \includegraphics[width=\textwidth]{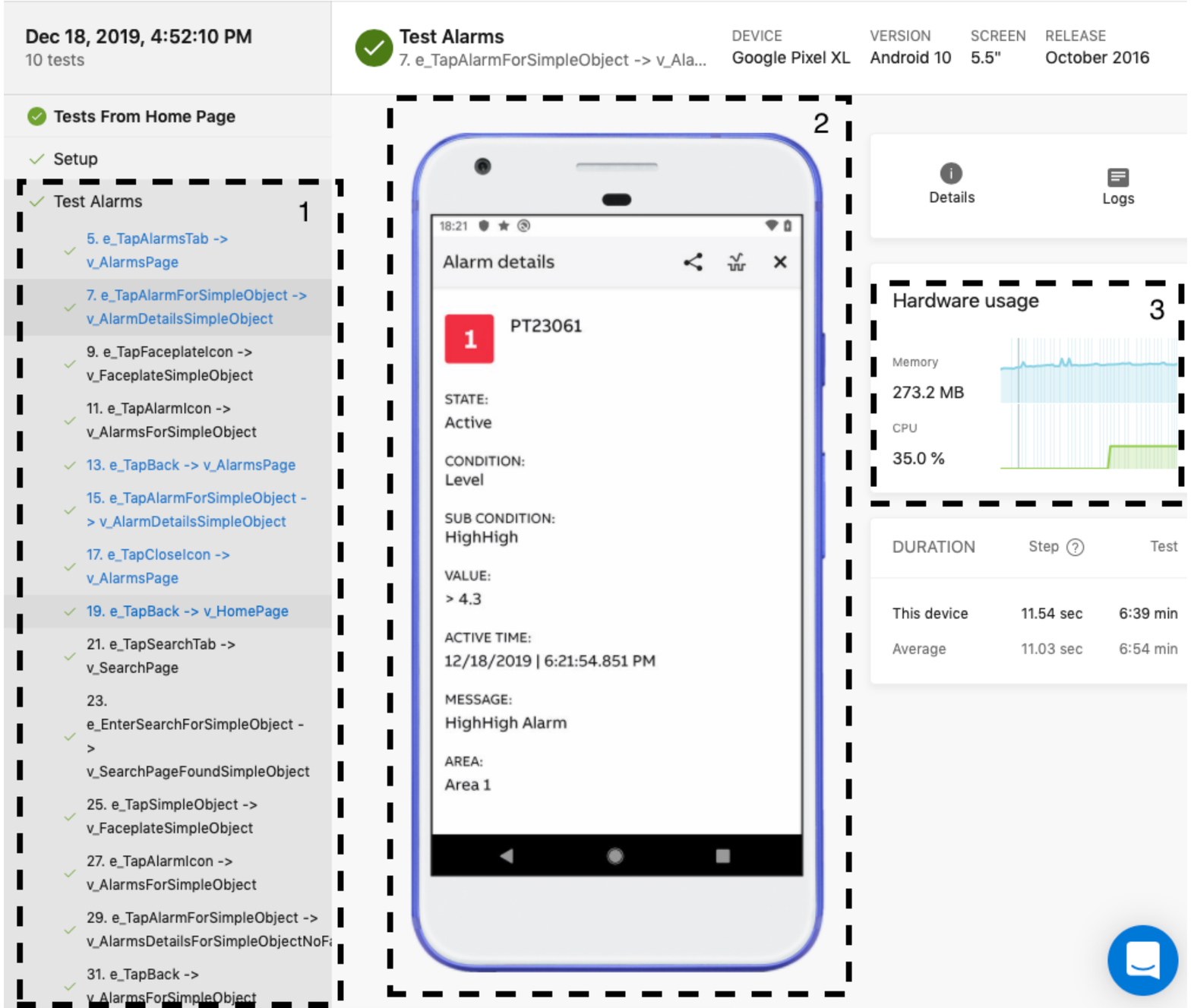}
    \caption{Overview of App Center.}
    \label{fig:app-center}
\end{figure*}

As described in our approach, an important property is to allow developers to execute tests on multiple devices. We have used the test capabilities of \textit{App Center} to enable this. By using one of the supported test frameworks, tests can be configured to run on multiple unique devices, with different OS versions and configurations. We created device sets containing devices with OS versions deemed as supported by our industry partner. In addition, given the capability of defining several different sets of devices, a specific set was defined containing the devices used by a pilot customer. This possibility gave the developers better confidence in the application developed since test execution covered all devices used in the planned pilot use case. 

A set of tests run on a set of devices and the recorded results are reported per device. Figure \ref{fig:app-center} shows an overview of the \textit{App Center} user interface for a specific test executed on a specific device. The left part (1) shows the test executed, including each step were a screenshot was recorded. By selecting a step in the list we can see a screenshot of how the GUI looked like (2) in that specific part of the test. In addition, the performance metrics of memory usage and CPU are shown to the right (3). By selecting a part of the metrics graph, the test step executed at that specific time is selected. This allows us to quickly find the step where any deviation of metrics started. Note, we relate every test step name (1) to its corresponding edge and node names, which is also used in the log files produced. This means that there is traceability from the model, through the test generation, implementation, and execution.

\section{Evaluation}

\begin{table}[t]
    \centering
    \caption{State transition coverage data. The Finite State Machine (FSM) column shows all the possible state transitions in the application, while the \textit{User Model} column shows the transitions related to the user interaction model.}
    \begin{tabular}{l |r r |r r}
        \hline
         & \multicolumn{2}{c }{FSM} & \multicolumn{2}{|c}{User Model}  \\
         \hline
         Covered & 92 & 47\% & 92 & 97\% \\
         Not Covered & 102 & 53\% & 3 & 3\% \\
         \hline
         Total & 194 & 100\% & 95 & 100\% \\
        \hline
    \end{tabular}
    
    \label{tab:coverage-result}
\end{table}{}

\begin{table}[h]
    \centering
    \caption{The distribution of reasons for lack of coverage.}
    \begin{tabular}{l|r|r}
        \hline
         Reason & Count & \% \\
         \hline
         Access Token & 46 & 39\% \\
         External Service & 44 & 37\% \\
         External Device Event & 15 & 13\% \\
         Discovered Implementation Error & 5 & 4\% \\
         Timing & 5 & 4\% \\
         User Model Coverage Miss & 3 & 3\% \\
         \hline
         Total & 118 & 100\%  \\
        \hline
    \end{tabular}
    
    \label{tab:coverage-reasons}
\end{table}{}

In this section we present an evaluation of the proposed approach in an industrial case study, considering the guidelines by Runesson et al.\cite{Runeson:2008:Guidelines-for-conducting-and-reporting-case-study-research}. Since this way of testing was new to our partner, evaluation of the usefulness and indications for further investment in a model-based approach to testing was of interest. Based on that, the research questions formulated to evaluate was:

\begin{itemize}
    \item \textbf{RQ1:} What coverage of functionality of the application under test is achieved with the used approach?
    \item \textbf{RQ2:} What are the limiting factors, if any, to achieve higher coverage of functionality?
\end{itemize}{}

With RQ1 we wanted to collect data to present to our industry partner, of the effectiveness of the used approach, to cover the functionality of the application under test. In addition, with a measurement of effectiveness, further evaluations of any improvement efforts needed could be performed.

By answering RQ2 we would, based on the coverage data from RQ1, then get a list of limiting factors as a basis to drive any further improvements of the used approach by our industry partner. Further, by answering RQ2, we would get an indication if there are areas lowering coverage more than others, providing information on where improvement efforts should be focused.

\subsection{Studied Case}
The application under test was a newly developed mobile application used by operators to view data from a distributed control system\footnote{\url{https://en.wikipedia.org/wiki/Distributed_control_system}}, such as process objects (motors, pumps, etc.), the current values produced by said objects (pressure, temperature, etc.) and any alarms produced. The application was developed using the Xamarin\footnote{\url{https://dotnet.microsoft.com/apps/xamarin}} framework, enabling cross-platform development, for Android and iOS, with the C\# programming language\footnote{\url{https://en.wikipedia.org/wiki/C_Sharp_(programming_language)}} on the .NET platform\footnote{\url{https://dotnet.microsoft.com/}}. The mobile application connects to a set of back-end services that provide the different kinds of data to be presented in the application. In addition, since cybersecurity is of high importance in the domain of control systems, the application's user must be authenticated, via a back-end identity proving service.

\subsection{Data Collection and Analysis}
Due to technical reasons for the Xamarin framework, bridging .NET and the Java runtime of an Android application, traditional tools for measuring code coverage can not be used. To still get useful coverage data we used a \textit{state transition} coverage metric for our data collection. The architecture of the application under test is based on a finite state machine (FSM). This allowed us to prepare the SUT for data collection by producing log file entries for each transition in the FSM. We then compared the performed transitions with all possible transitions, as defined in the source code, resulting in a coverage metric of the implementation. This method of measuring coverage produces a metric on a higher abstraction level then code-coverage but lower than on a feature level. The implementation described in Section \ref{section:implementation}, where tests were generated from a user-based model and executed on App Center, was applied to the mobile application under test to collect the state transition coverage data.

The analysis was performed based on the collected coverage data. We went through each non-covered state transition with the responsible developers, categorizing the reason for each transition not to be covered, with respect to limitations in the test approach applied.

\subsection{Results}
In this section, we present the result of the performed evaluation and the implications of those results.

\subsubsection{RQ1: Coverage}
As can be seen in Tabel ~\ref{tab:coverage-result}, The total \textit{state transition coverage} of the FSM was 47\%. It is unsurprising that an application that relies on an identity provider and external services are not completely covered by a model modeling a user, but prior to actually measuring, there was no knowledge of how large part of the SUT that was associated with such functionality, and thus not covered. Reviewing coverage data, analyzing which transitions that could be covered by the existing model or any improvements, from a user-model perspective to it, we ended up with coverage of 97\%. Thus, an improved user-based model could only increase the overall coverage of the application by a small percentage, any significant improvement would have to include other considerations than the user. For example, no improvement in the user-based model would cover states related to connection issues with external services.

\begin{result}
The coverage of the applied approach, from a user-model perspective, is high (97\%) but the overall coverage of the application under test (47\%) leaves room for improvements.
\end{result}

\subsubsection{RQ2: Limiting Reasons for Coverage}
As can be seen in Table ~\ref{tab:coverage-result}, the FSM contained 194 transitions in total with 102 not covered. Analyzing the reasons for each of those we identified six different categories of reasons why a transition was not covered. A summary of the reasons is shown in Table ~\ref{tab:coverage-reasons}. Note that the total number of reasons (118) in Table \ref{tab:coverage-reasons} is higher than the not covered transitions (102) in Table \ref{tab:coverage-result}, this is due to that to cover a single transition multiple conditions might have to be fulfilled, resulting in a single transition counting towards multiple reasons.

The six identified reasons that limit the transition coverage were:
\begin{itemize}
    \item \textbf{Access Token} - This includes transitions that depend on the state of the applications provided access token, such as expiration date and refresh status.
    \item \textbf{External Service} - These transitions depends on events of external back-end services such as loss of connection, re-connections, etc.
    \item \textbf{External Device Event} - This reason includes events external to the application under test happening on the device. For example, a phone call is received or the user context switches to another application, putting the application under test in the background.
    \item \textbf{Discovered Implementation Error} - While analyzing the transitions not covered, the developers found transitions that were deemed as implementation errors such as transitions not longer relevant, i.e., dead code.
    \item \textbf{Timing} - Reasons depending on concurrent events such as the order in which data is received while navigation changes occur.
    \item \textbf{User Model Coverage Miss} - This category covered the reasons were there was a lack in the used model, i.e., an improved model would cover these transitions.
\end{itemize}{}

\textit{Access Token} (39\%) and \textit{External Service} (37\%) was the dominant reasons, totaling 76\%, of the non-covered transitions. This makes those reasons the main candidates for evaluating future improvements to the implemented approach.

\begin{result}
The primary limiting factors to achieving a higher coverage were due to access token status and the communication with external services, indicating that to achieve a high coverage the used approach must be expanded to include the possibility for test generation of these limiting factors.
\end{result}{}

\section{Lessons Learned}

In this section, we describe the lessons learned collected during, the more than twelve months of, the project presented in this paper. The lessons relate both to the approach and its implementation, but also to the lessons of transferring knowledge from academia to industry. 

\textbf{Lesson 1: Seeing is believing.} We attribute a large part of the success of this project and industry-academia knowledge transfer to the idiom \textit{"Seeing is believing"}. We started out with a small proof-of-concept that was introduced to a small number of developers, introducing the technique of model-based testing, which was new to the developers. We believe, and the developers confirm this, that just lecturing the benefits and capabilities of model-based testing would not have had the same effect as an actual running example. This was the first step of \textit{"Seeing is believing"}, which was enough for one development team to be interested and adopting the approach described in this paper.

The second stage was done when the adopting team had built and end-to-end implementation of the approach, as described, and made an in-detail demonstration to a larger group of multiple development teams. The presentation was done by one of the developers involved in making the implementation, using the real application as the SUT. This delivered high credibility of the approach of being useful in practice. The effect of the demonstration was that another team started to use the same approach to generate tests for their web application.

This process of transfer of the approach within the organization continuous, spreading further from the original team, and more teams are requesting demonstrations.

The learning we want to express is the power of running examples applied in the context of the target practitioners, and not only to rely on theoretical models and publications.

\textbf{Lesson 2: Practical Tooling.} Practitioners are usually pressed for time and, in our experience, testing activities related to test automation and test generation will lose in scheduling competition with user-facing features. With this in mind, tools used for test automation and generation must be easy to pick-up and get started with by practitioners. We believe we found a good balance between simple and easy to use off-the-shelf tools.
Further, we believe that by using simple and practical tooling and actually getting the project implemented and model-based testing adapted, we paved the way for future knowledge transfers. With an end-to-end test generation process in place, not perfect but useful, we believe it to be far easier to transfer incremental improvements than starting new projects requiring a larger time investment.

\textbf{Lesson 3: Benefits of modeling.} In addition, to be used for test generation, the model was also used as a basis for design discussions in the development teams. This provided added benefits, several problems were realized and solved while modeling the applications' interaction model instead of at a later, and probably more costly, stage of the development cycle. By using a model on a high abstraction level the model was thus reusable on both the Android and iOS versions of the mobile application developed, reducing the amount of work to test for different platforms. These benefits increased the perceived value of developers of spending time creating the models required for test generation.

\textbf{Lesson 4: Implementation Simplicity}
Be keeping the node and edge concept from the model throughout to the implementation of the test runner, we keep the complete process simple, transparent and easy to follow. Failing tests and log output can easily be traced back to the model and the path taken, reducing the effort to reproduce the failing scenario. In addition, this method has kept the maintenance effort of the tests low, since changes in the application can be reflected in the model and node and edge implementations already in place are automatically reused. Also, any change in the behavior of performing a transition action or asserting in the GUI is isolated to a single node or edge implementation, which is highly reused for common edges, further simplifying maintenance.

\textbf{Lesson 5: Build a knowledge transfer pipeline}
To increase the likeliness of a successful knowledge transfer we learned the value of having a transfer "pipeline", where information can flow, spanning from both the poles of academia and industry but also including people connected to both environments. Our setup, which worked very well, consisted of flowing information, in both directions, from senior academics in software testing via an Industry Ph.D. student, via a champion in industry and to a development team. The champion is an individual who is interested in the topic of the transfer and who can represent the project in planning activities in industry. In addition, it is preferred if this individual have the capacity to make the proof-of-concept implementations in the real industry context, used to deliver the "seeing is believing" experience to the rest of the organization.  

\textbf{Lesson 6: Usefulness of Randomness}
As described, we used a random walk of the feature under test, randomly selecting edges to be traversed in the model. This turned out to be a useful strategy that found real bugs due to producing sequences in the model not considered by the developers. This strategy also increased the perceived usefulness of the test generation process by the developers, since they perceived more value from randomly creating new test cases than, the more commonly used strategy, of only executing static test cases.

\section{Related Work}

In a systematic literature review, including 103 publications until 2016, Kong et al. identify the leading methods for Android application testing. Model-based testing (65), black-box methods (52), where no knowledge of the SUT is required, system testing (81),  were the whole app is tested, and deployed on real devices (68), was the most applied methods \cite{Kong-Automated-Testing-of-Android-Apps-review}. Considering the targets and objectives of publications, Kong et al. also identified that the issues of compatibility, i.e., the set of devices an application can execute on without any problems, is understudied \cite{Kong-Automated-Testing-of-Android-Apps-review}. Our applied approach maps to the methods most commonly used, but in addition, we contribute to the compatibility gap, since we execute on multiple real devices hosted in the cloud.

Recently, the testing of compatibility issues of Android mobile applications have gained some attention. Ki et al. propose \textit{Mimic}, a tool for UI compatibility testing. Mimic provides a runtime where an application can be tested on several different OS versions and devices, using several testing strategies, such as comparing the visual layout of the application ~\cite{Mimic}. This is an interesting approach, but we have opted for an off-the-shelf solution to handle the different sets of devices and their configurations. While not as powerful as Mimic, lacking the ability for automatic visual comparison, but with the aim of industry adoption and knowledge transfer, the availability and possibility of integration in the workflow of practitioners have been critical in the selection of our approach.

A common theme of works in model-based testing of mobile applications has been the automatic creation of the model ~\cite{Sapienz, MobiGUITAR, Multi-Level-GUI-Comparison-Criteria, Guided-Stochastic-Model-Based-GUI-Testing-of-Android-Apps, Guided-GUI-Testing-of-Android-Apps, Practical-GUI-Testing-of-Android-Applications, AimDroid, A-Grey-Box-Approach-for-Automated-GUI-Model-Generation}. Automatically creating models can greatly improve the generalizability of an approach, however, there is a trade-off in the types of faults that can be found with the lack of application-specific behavior included in the model. In general, we differentiate in that we use a human-in-the-loop, to create the interaction model of the application under test, making sure the expected behavior of the application under test is included. 

MobiGUITAR is an automated model-based testing tool for mobile applications. MobiGUITAR builds the GUI model of an application by traversing the applications GUI, with the result of a state machine model. The tests generated are based on coverage of the model, where pairs of edges adjacent to a node in the model should be covered \cite{MobiGUITAR}. Since the model is based on the GUI, MobiGUITAR will not find behavior problems, any navigation bugs introduced in the GUI will be present in the model. Our method is based on a human-made model to ensure the correct behavior. In addition, our approach generates random walks in the model, increasing the possibility of finding problems that require several stateful interactions.

SAPIENZ uses a fully automated multi-objective search-based approach to Android application testing by optimizing for code coverage, the length of a test sequence and fault finding in the form of crashes \cite{Sapienz}. As with MobiGUITAR, since there is no behavioral model, the tool can find crashes but not verify behavior. In addition, the efforts of SAPIENZ resulted in a startup that was later acquired by Facebook \cite{Kong-Automated-Testing-of-Android-Apps-review} and the code repository\footnote{\url{https://github.com/Rhapsod/sapienz}} is no longer maintained, making it less attractive to apply in an industry context.

Baek et al. propose a method that, given different levels of GUI comparison criteria, automatically creates a model of the GUIs different distinct screens by dynamically generating events and inputs, building the model in an iterative fashion, given the changes in the GUI \cite{Multi-Level-GUI-Comparison-Criteria}. We do not use a detailed model of the user interface, instead, we model at a higher level of abstraction of the perspective of the users' actions and user-perceived states, allowing the model to also be a design and documentation artifact, in addition to test generation, that are more related to the requirements.

\textit{Stoat} improves the fault-finding effectiveness of MobiGUITAR\cite{MobiGUITAR} and SAPIENZ\cite{Sapienz} by using a stochastic dynamic modeling process. In addition, Stoat randomly injects system-level events in the generated tests, such as screen orientation changes, phone calls, and network connection changes, to improve the fault-finding capabilities of the method. This injection of events does not increase the complexity of the model, by including system events into the behavior model \cite{Guided-Stochastic-Model-Based-GUI-Testing-of-Android-Apps}. Currently, our model contains the system-level events of installing and uninstalling the app, but the inclusion of other system-level events would be an interesting addition to further strengthen the test suite.

A more recent, and also the most effective, compared to SAPIENZ and Stoat, fully automatic model-based approach is APE. APE uses a dynamic modeling approach, where the abstraction granularity of the model is adjusted to the application under test during testing \cite{Practical-GUI-Testing-of-Android-Applications}. The value of using different abstraction levels for a manually created model, as in our case, could be an interesting avenue to investigate for future work. Such an approach could allow the model to both stay close to the requirements and features of the application, but where needed, also go into finer details of GUI behavior.

Takala et al. report similar experiences to us. Applying a model-based approach, they found bugs both while modeling the SUT and while executing generated tests. In addition, they found the maintenance effort to bee low for their model-based tests. Further, as we also did, they experienced the advantages of random generation of test cases to produce sequences that designed tests would not do. However, their tooling was perceived as requiring significant time investment to use by practitioners \cite{experiances-of-system-level-MBT-GUI-testing}. The ease of usage is an area which we have aimed to improve, in order to make a model-based testing approach accessible to practitioners.

\section{Conclusion}
In this paper we have presented an approach of applying model-based testing to testing a real-world mobile application, executed on multiple real devices, hosted in the cloud. We have also reported on the lessons learned of performing this project together with an industry partner, transferring knowledge from academia into a real running test generation process. It turns out that this is a useful approach that, with the right guidance, can be successfully transferred and used in industry. Our evaluation shows promising results of covering the functionality of the SUT with room for improvements, by expanding the approach to include external services and authorization into the test generation process.

Our main contribution to practitioners is the description of the approach used and a concrete implementation example that can be adapted to other practitioners' contexts. For researchers, we believe that our lessons learned can help and inspire future knowledge transfer of research results in academia to industry.

For future work, we are now in a position at our industry partner where we can make improvements to the already-in-place, thanks to this project, model-based test generation process. Such an improvement would be to include a larger part of the real industry system into the test generation, such as the back-end service used by the mobile application. Such a system-level approach would allow for test generation to be more useful in system-level black-box testing, an area which is now tested with high manual effort. 

\section*{Acknowledgements}
The authors would like to thank Ola Ottemalm for valuable feedback.
This work is supported by ABB, the industrial postgraduate school Automation Region Research Academy (ARRAY) funded by The Knowledge Foundation. Additional support is provided by ITEA3 TESTOMAT project funded by VINNOVA.

\bibliographystyle{IEEEtran}
\bibliography{main}
\end{document}